\documentclass[12pt]{article}

\usepackage{scicite}
\usepackage{graphicx}

\usepackage{times}

\newenvironment{sciabstract}{%
\begin{quote} \bf}
{\end{quote}}

\newcounter{lastnote}

\title{Digital-photonic synthesis of ultra-low noise tunable signals from RF to 100 GHz}

\author
{T.M. Fortier,$^{1\ast}$ A. Rolland,$^{1}$ F. Quinlan,$^{1}$ F.N. Baynes,$^{1}$\\ A.J. Metcalf,$^{2}$ A. Hati,$^{1}$ A.D. Ludlow,$^{1}$ N. Hinkley,$^{1}$ M. Shimizu,$^{3}$ T. Ishibashi,$^{3}$\\ J.C. Campbell$^{4}$ and S.A. Diddams$^{1\ast}$   \\
\\
\normalsize{$^{1}$NIST, Time and Frequency Division, 325 Broadway MS 847, Boulder CO 80305, USA}\\
\normalsize{$^{2}$Purdue University, West Lafayette, IN 47907, USA}\\
\normalsize{$^{3}$NTT Electronics Corporation Naka-shi, Ibaraki-ken, 311-0122, Japan}\\
\normalsize{$^{4}$University of Virginia, 351 McCormick Road, Charlottesville, VA, USA}\\
\\
\normalsize{$^\ast$To whom correspondence should be addressed; 
E-mail:  fortier@boulder.nist.gov}\\
\\
\normalsize{A. Rolland and F.N. Baynes are now with National Physical Laboratory, Teddington, UK}
}

\date{}

\begin{document} 


\baselineskip24pt


\maketitle


\begin{sciabstract}
The demand for higher data rates and better synchronization in communication and navigation systems necessitates the development of new wideband and tunable sources with noise performance exceeding that provided by traditional oscillators and synthesizers. Precision synthesis is paramount for providing frequency references and timing in a broad range of applications including next-generation telecommunications, high precision measurement, and radar and sensing. Here we describe a digital-photonic synthesizer (DPS) based on optical frequency division that enables the generation of widely tunable signals from near DC to 100 GHz with a fractional frequency instability of 1 part in 10$^{15}$. The spectral purity of the DPS derived signals represents an improvement in close-to-carrier noise performance over the current state-of-the-art of nearly 7 orders of magnitude in the W-band (100~GHz), and up to 5 orders of magnitude in the X-band (10~GHz). 
 \end{sciabstract}

\section*{Introduction}
In standard synthesizer architectures, the close-to-carrier noise of the synthesized signal is determined by the multiplied noise of a quartz crystal reference. Recent efforts in photonic~\cite{capmany2007microwave,khan2010ultrabroad,maleki2011sources,fortier2011generation,li2014electro} and MEMS~\cite{nabki2009highly,lee2012low,phan2013high} based sources have been driven by the prospect for devices with better timing and higher carrier frequency than crystal oscillators. Photonic approaches to signal generation offer wideband operation at sub-THz frequencies with fast tuning and high signal-to-noise ratio~\cite{song2007generation,song2008broadband,Jones:14,rolland2014narrow}.  However, because many photonic architectures still rely on quartz oscillators as a reference, their noise performance is not significantly improved over what can be achieved with conventional electronics~\cite{schneider2013radiofrequency,pillet2014dual}. 

A synthesizer using a time-base derived via optical frequency division (OFD)~\cite{hati2013state} offers several technical advantages over techniques based on electronic crystal oscillators (See Fig.~\ref{figure1}). Optical reference cavities exhibit extremely low loss and drift and thereby achieve much higher quality factors (Q$\sim$10$^{11}$) and lower instabilities ($<$ 10$^{-15}$ at 1~s) than electronic resonators~\cite{jiang2011making,kessler2012sub}. Synthesis begins with the reference at optical frequencies (100’s of THz), yielding reduction in the reference noise power on a derived microwave carrier by the division factor squared. High power and high linearity photodiodes now allow for generation of Watt-level microwave carriers~\cite{li2011high,xie2014improved}, and residual noise supporting frequency instabilities of 10$^{-17}$ at 1 s averaging on a 10 GHz carrier~\cite{Baynes:15}. Finally, OFD produces a microwave harmonic spectrum of equally spaced tones, from DC up to the photodiode bandwidth (e.g. typically 10-100 GHz), each carrying the stability of the optical reference and with a noise floor approaching 18 orders of magnitude below the carrier~\cite{quinlan2013exploiting}. 

While OFD provides a set of high spectral purity reference frequencies, ultimately the broadest utility requires transfer of the optical stability to any radio, microwave or millimeter-wave frequency in a user defined, frequency-agile way. For example, frequency hopping communications~\cite{vig1992military}, as well as stepped and continuously swept frequency radar require frequency agility and fast tuning while maintaining high spectral purity~\cite{scheer1993coherent,ghelfi2014fully}. Here we accomplish broadband frequency agility, precision tuning, and ultra-low phase noise simultaneously by using OFD as the timebase to a digital-photonic synthesizer (DPS).

\section*{Results}
\subsection*{Photonic timebase}

\begin{figure}
\centering
\includegraphics[width=12cm]{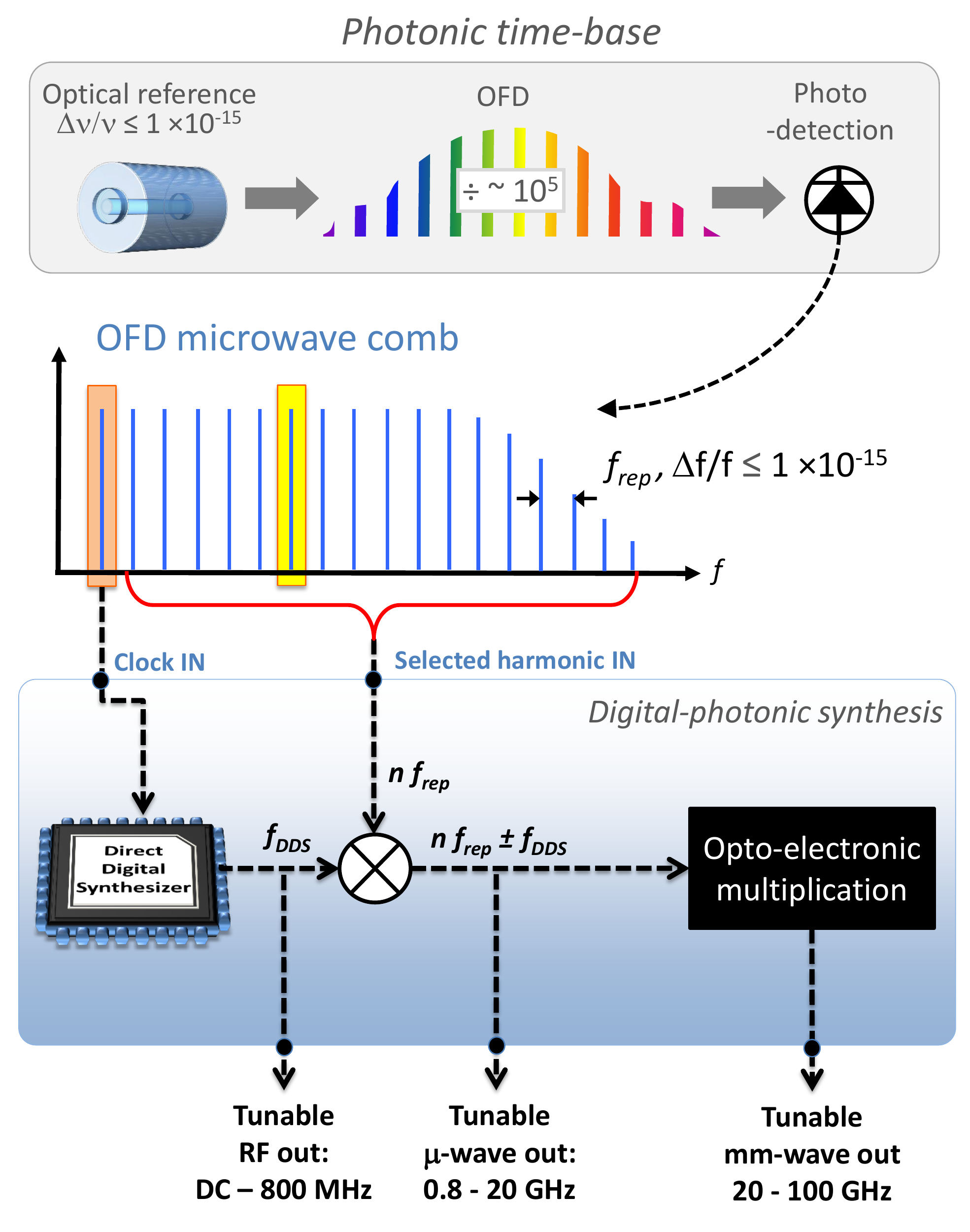}
\caption{The simplified and generalized architecture for a digital-photonic synthesizer (DPS). A low drift and high stability optical cavity is used as a reference to stabilize the spectrum of an optical frequency comb divider (OFD). Photodetection of the stabilized OFD spectrum results in a microwave harmonic comb spanning the photodetector bandwidth with a spacing of $f_{rep}$. This microwave harmonic comb serves as the time base to a digital synthesizer. In our particular implementation, the lowest order $f_{rep} = $~2~GHz harmonic is used as a clock from which a direct digital synthesizer (DDS) digitally synthesizes phase tunable RF frequencies from $f_{DDS} =$ DC to 800 MHz.  This tuning range is extended up to 20~GHz by mixing the output of the DDS with subsequent harmonics, $nf_{rep}$, of the microwave harmonic comb. Opto-electronic multiplication by 10 of ($nf_{rep} + f_{DDS}$) allows for extension of synthesis to mm-wave frequencies (20$-$100~GHz).}
\label{figure1}
\end{figure}

Fig.~\ref{figure1} shows a simplified experimental setup for our DPS. A more detailed explanation of our procedures can be found in the Methods section. The timebase for our synthesis technique is a microwave frequency comb derived via division by ~10$^5$ of an ultra-stable frequency reference at 518~THz. Optical frequency division results in a microwave comb and not a single microwave frequency because the optical divider itself is an optical frequency comb based on a mode-locked laser. Stabilizing the optical frequency comb to the optical reference transfers its stability to the timing in the mode-locked optical pulse train~\cite{ramond2002phase, fortier2011generation, fortier2013photonic}. Photodetection converts the stable optical pulse train to a stable electronic pulse train whose frequency spectrum comprises discrete and equally spaced tones, i.e. the microwave frequency comb. The lowest order harmonic represents the pulse repetition rate, $f_{rep}$, which is 2~GHz for our experimental demonstration. Any one of the myriad tones generated within the bandwidth of the photodiode (PD) may be used as a low noise electronic source.

\subsection*{Radio frequency digital-photonic synthesis}

Some frequency agility of the microwave comb is possible by tuning the repetition rate of the optical pulse train, but this tends to be slow and limited to 1-2\% due to stability constraints of the mode-locked laser cavity~\cite{washburn2004fiber}. A combination of OFD and regenerative frequency division can yield RF signals with extremely low noise~\cite{hati2013state}, but this does not enable frequency tuning. Direct digital synthesis (DDS) offers an attractive approach to synthesis because with a single clock input it allows for digitally generated signals with fast slew rates ($< 1$ $\mu$s), precise phase tuning and low residual  noise~\cite{calosso2012phase}.

The DDS on its own exhibits very low residual noise that is typically overwhelmed by the timing error of a quartz based clock signal. The 2~GHz clock supplied via OFD provides a timing stability that is 100 times better than oven-controlled quartz, and 20 times better than BVA quartz~\cite{chauvin2007new}. As a result, using OFD as the clock to a DDS we can produce a continuously tunable RF source whose noise is limited only by the digital electronics, yielding signals with noise that is lower than a state-of-the-art quartz-based synthesizer.

\begin{figure}
\centering
\includegraphics[width=11cm]{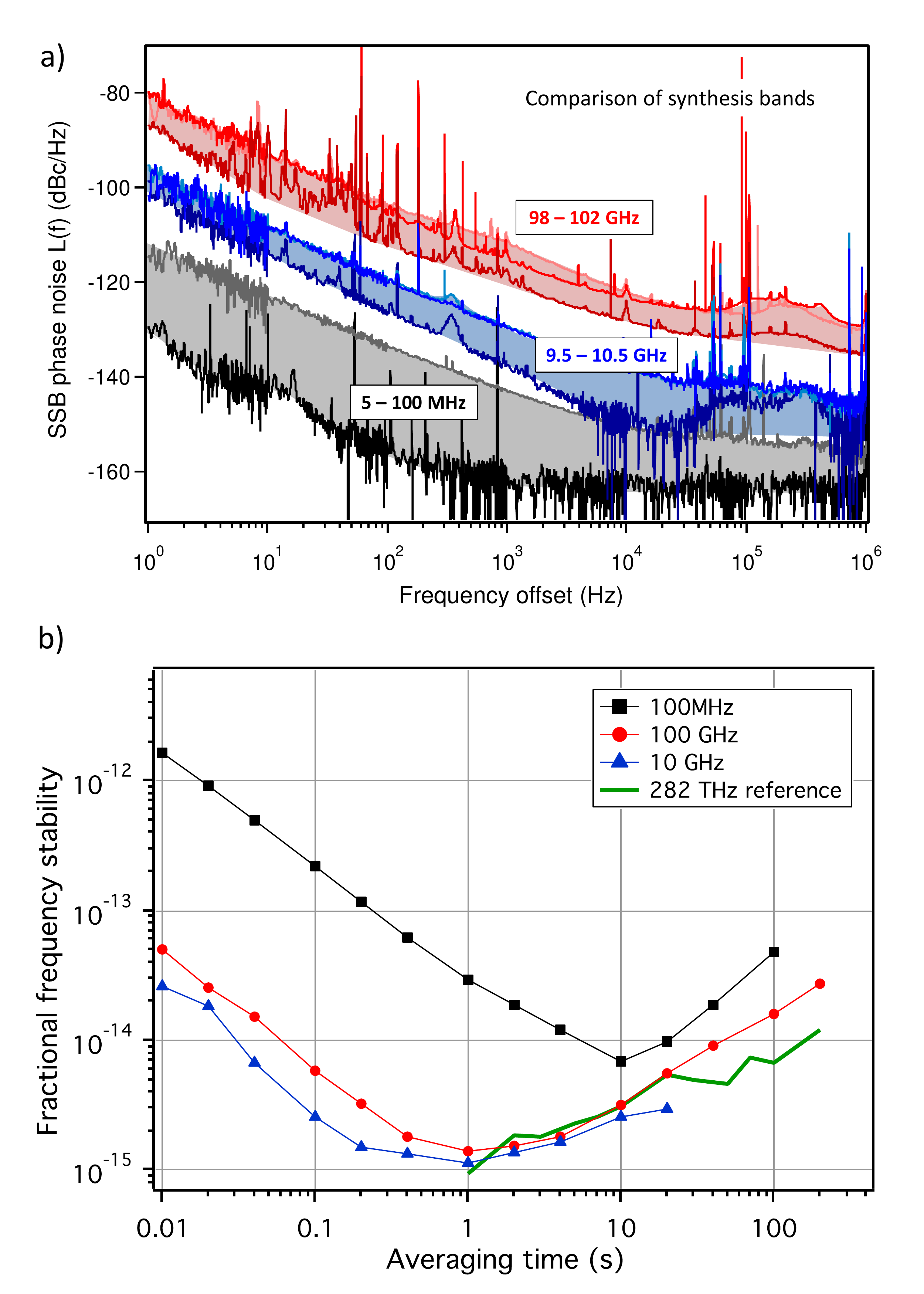}
\caption{The two oscillator comparison for characterization of the a) phase noise and b) fractional frequency instability of signals derived via photonic-digital synthesis in three frequency bands: RF (black), X-band (blue), and W-band (red). In a), the single sideband phase noise measures the level of the noise sidebands relative to the stated carrier frequencies.  For the signals synthesized from $5-100$ MHz, the noise is lowest at 5 MHz and increases until the top of the band at 100 MHz. For synthesized signals about 10 and 100 GHz, the noise increases with tuning away from the unshifted harmonics at 10 and 100 GHz. Also included in b) is the instability of the optical reference at 282 THz, which limits the frequency instability of the 10 and 100 GHz synthesized signals at times greater than 1~s.}
\label{figure2}
\end{figure}

Fig.~\ref{figure2}(a) shows the absolute phase noise of signals synthesized over a frequency band from 5$-$100 MHz by clocking a DDS with the lowest order OFD microwave harmonic at 2~GHz. The absolute phase noise was measured via comparison to a second independently clocked DDS, comprising a second OFD referenced to an ultra-stable optical cavity at 282~THz. Although the DDS can digitally synthesize signals up to 40\% of the clock frequency, the displayed frequency range in Fig.~\ref{figure2}~(a) from 5$-$100~MHz was chosen to allow for a clear distinction in the noise between the DPS-RF, DPS-X-band and DPS-W-band signals. We find that the baseline of our measurements is limited by the residual noise of the DDS. Nonetheless, the signals derived via the OFD driven DDS  yield tunable signals with a fractional frequency instability of less than 3 parts in $10^{14}$ at 1~s averaging (see Fig.~\ref{figure2}(b)), and exhibit phase noise levels comparable or better than that achieved by state of the art, fixed-frequency crystal quartz oscillators.

\subsection*{X-band digital-photonic synthesis}
Currently the maximum clock frequency of a state-of-the-art DDS is a few GHz, limiting its tuning range. However, using simple mixing techniques, we can leverage the microwave comb to extend the synthesis range by using the RF signal from the DDS to up- or down-shift the frequency of any harmonic within the OFD spectrum. In principle, shifting an OFD harmonic with the output of the DDS by up to $\pm f_{rep}$~/2 would allow for the generation of any frequency from near DC up to the highest harmonic detected from the OFD (up to 20 GHz with the photodiode used in our experiments here). 

Although the above described synthesis scheme could allow for continuous tuning from near DC to 20 GHz, in the measurement below we only explore synthesis in a limited band around 10 GHz since agility at this frequency has application to pulsed X-band radar~\cite{scheer1993coherent,fontana2004recent}.

As seen in the lower portion of Fig.~\ref{figure1}, generation of the agile 10~GHz signal is obtained by combining the 5th harmonic at 10 GHz from the OFD harmonic spectrum with the synthesized output from the DDS using a single sideband mixer~\cite{nand2011ultra}. The single sideband mixer allows for efficient up- or down-shifting of the OFD harmonic frequency while suppressing the local oscillator and image signals. Subsequent comparison with a second nearly identical but independent system allows for characterization of the tunable 10~GHz signal phase noise at times less than 1~s, as well as characterization of the signal stability at times greater than 1~s.

Fig.~\ref{figure2} (a) and (b) show the results of the absolute phase noise and fractional frequency instability measurements of the synthesized signals from 9.5 GHz to 10.5 GHz. In Fig.~\ref{figure2} we see that the phase noise of the native 10 GHz harmonic is -105~dBc/Hz at 1 Hz offset. Synthesis about this carrier results in added noise from the DDS. For offsets from the 10 GHz harmonics greater than 250 MHz, the added noise of the DDS dominates that of the unshifted 10 GHz signal. Nonetheless, in Fig.~\ref{figure2} we observe that the synthesized signals from 9.5~GHz to 10.5~GHz maintain a \textit{combined} close-to-carrier noise below -97~dBc/Hz at a 1~Hz offset and  fractional frequency stability below $2\times10^{-15}$ at 1~s averaging time.

\begin{figure}
\centering
\includegraphics[width=12cm]{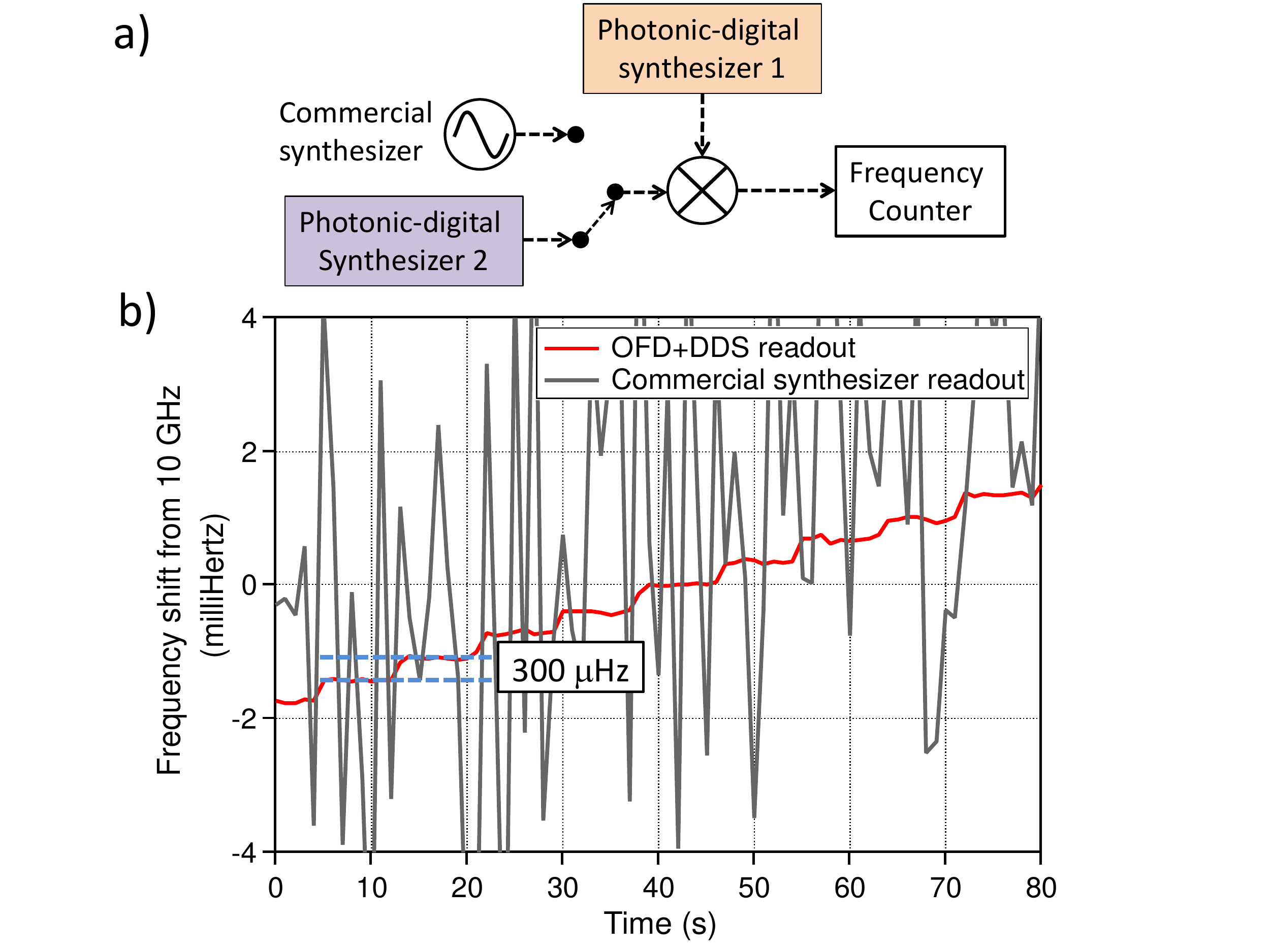}
\caption{a) Simplified setup for electronic readout of a finely stepped DPS near 10~GHz with, in one case, a H-maser referenced synthesizer and, in the other case, with a second 10~GHz DPS source. b) Readout of a stepped 10~GHz DPS source with (grey line) a H-maser referenced synthesizer and (red line) a second 10~GHz DPS source counted with  a 1~s gate time.}
\label{figure3}
\end{figure}

A source that combines low noise with fine frequency resolution is important in improving the security in communications by enabling both faster carrier detection and frequency hopping~\cite{vig1992military}. As a test of the resolution possible with our agile signals we demonstrate fine frequency stepping of the 10 GHz source. Figure~\ref{figure3}~(a) shows the simplified setup for the readout of a stepped 10 GHz signal from a photonic-digital synthesizer. In our measurement the step size was 300~$\mu$Hz. The 10~GHz stepped source from one digital-photonic synthesizer was mixed with a local oscillator near 10~GHz and the down-converted signal was counted with a gate time of 1~s. Figure~\ref{figure3}~(b) shows the readout of the stepped signal as performed with two local oscillators; a fixed 10 GHz signal from a H-maser referenced synthesizer, and a fixed 10~GHz signal from a second DPS. As seen in Fig.~\ref{figure3}~(b), it is only possible to resolve the fine frequency steps of the photonically generated 10~GHz signal with a second DPS since the electronically referenced synthesizer lacks the necessary frequency resolution. 

\subsection*{W-band digital-photonic synthesis}
Extension of digital-photonic synthesis beyond the bandwidth of the photodetector can be achieved via several different architectures. Electronic multiplication can be employed to multiply X-band signals to the W-band~\cite{hati2014,bara2012}. While this method is adequate for quartz-based synthesizers, it degrades the spectral purity and limits the high-frequency noise floor (as will be discussed in greater detail below) of our DPS signals.  An additional drawback of electronic multiplication is that it does not allow for efficient distribution of W-band signals, which would require expensive and lossy waveguide components. In our approach below we employ opto-electronic  multiplication of X-band signals derived via DPS to enable lower-noise synthesis and efficient fiber-optic distribution of ultra-low noise  mm-wave signals.

\begin{figure}
\centering
\includegraphics[width=10cm]{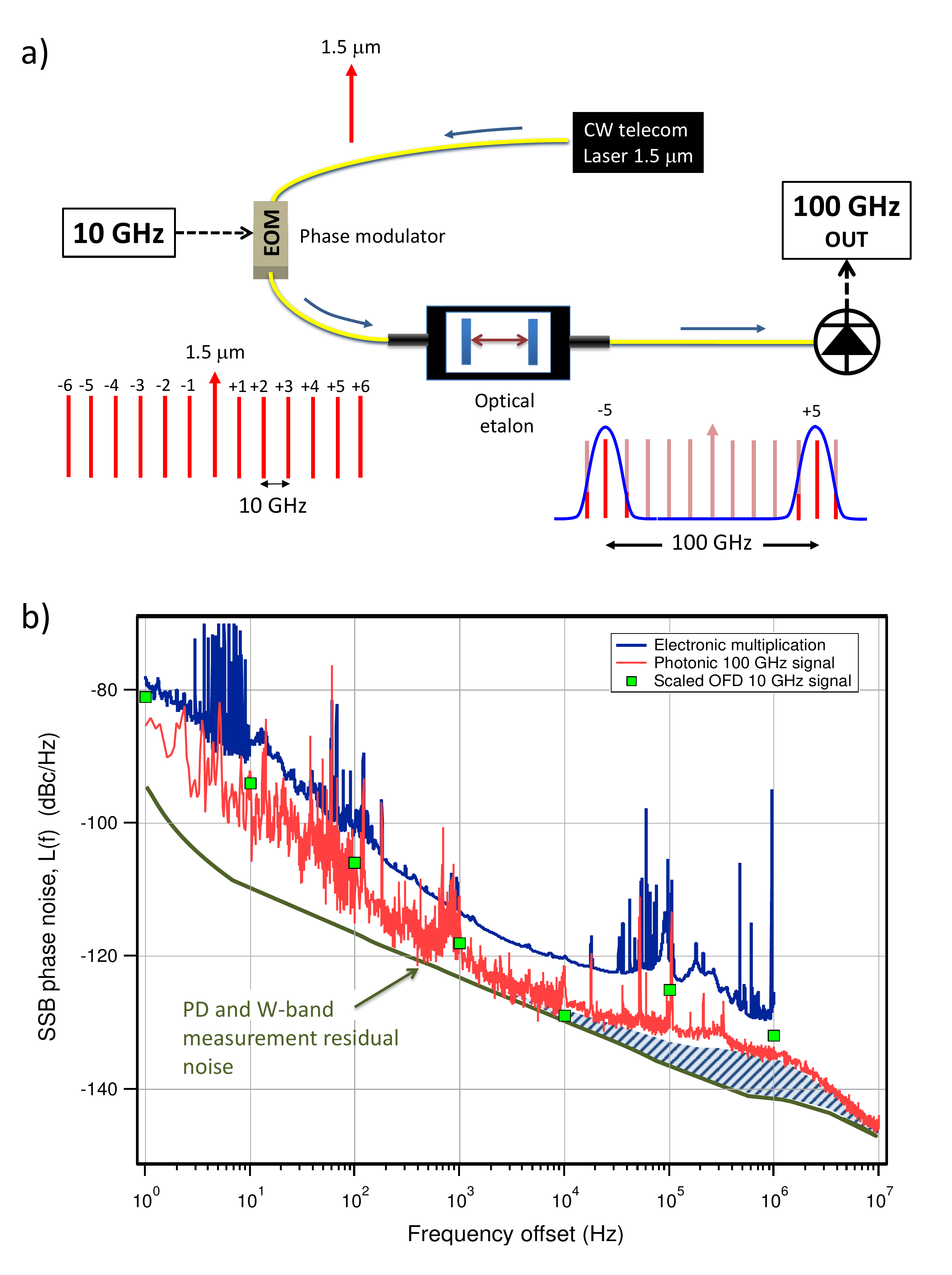}
\caption{a) Simplified experimental setup for opto-electronic multiplication of 10~GHz signals to 100~GHz.  Sidebands spaced by 10 GHz from the X-band DPS  are modulated onto a 1.5~$\mu$m laser with an electro-optic modulator (EOM) driven by the 10 GHz from the DPS. An optical etalon is used to filter two 100~GHz spaced optical modes for efficient photodetection of a 100~GHz electrical signal. b) Characterization of noise sources at 100~GHz. Absolute phase noise of two 90~GHz signal obtained via 9 times electronic multiplication (blue trace) and two 100~GHz signals obtained via opto-electronic multiplication (red trace) of photonically generated 10~GHz signals. The green trace shows the baseline level obtained for measurement of the residual noise of the 100~GHz photodetection. The blue shaded area shows the excess noise in opto-electronic multiplication. The filled green squares shows the calculated 10 times multiplication of the 10~GHz phase modulator drive signal.}
\label{figure4}
\end{figure}

As seen in the lower right inset in Fig.~\ref{figure4}(a), we achieve 10 times opto-electronic multiplication of our 10~GHz DPS signal by phase modulating a continuous wave (CW) laser at 1550 nm with an electro-optic modulator (EOM)~\cite{o1992optical,song2007generation,song2008broadband,hirata2003120,li2014photonic}. The EOM is overdriven by the agile 10 GHz signal producing phase modulation sidebands (up to greater than $\pm 5$) about the CW optical carrier.  The resulting 10 GHz spaced EOM comb was then centered midway between the modes of a 100~GHz spaced optical etalon. Filtering with the optical etalon allowed for selection of two optical modes separated by 100 GHz with an average optical power of 15~mW. Subsequent photodetection with a waveguide-coupled unitraveling-carrier (UTC) photodiode~\cite{ito2004high,ishibashi2015ultrafast} allowed for derivation of an electrical 100~GHz signal with $-5$~dBm carrier strength. 

We derived tunable mm-wave signals from $98-102$ GHz by driving the EOM with agile 10 GHz signals. The frequency band was limited only by the 100 GHz etalon whose optical modes exhibit only a few GHz of bandwidth. Agility over a larger band would be possible with a tunable optical etalon or a reconfigurable optical filter. 

Characterization of the millimeter-wave spectral purity was obtained via comparison to a second independent system comprising a second W-band DPS. Fig.~\ref{figure2} shows the measured phase noise and fractional frequency instability of the generated W-band signals. As seen in Fig.~\ref{figure2}~(a), noise added by the DDS on the agile 10 GHz source resulted in a slightly higher phase noise at all offset frequencies. For all synthesized signals in the band 98$-$102 GHz, the absolute phase noise was less than -80 dBc/Hz at 1 Hz offset, falling to a noise floor of -130 dBc/Hz at 1 MHz offset from the carrier. The corresponding integrated timing noise was less than 1 fs integrated from 1 Hz to 1 MHz.  Also seen in Fig.~\ref{figure2}~(b), the fractional frequency instability was less than $2 \times 10^{-15}$ at 1~s averaging time, reproducing the stability of the optical reference cavities.

Fig.~\ref{figure4}~(b) shows a comparison between the absolute phase noise of W-band signals generated via optical and electronic multiplication of 10 GHz signals generated via DPS.  Also shown is the residual noise contributed by the opto-electronic multiplication setup, including the 100 GHz phase noise measurement setup. The dominant source of noise for offset frequencies from 1~Hz$-$1~kHz results from opto-electronic conversion in the high speed photodetectors (green trace), from which we can estimate an upper limit to the flicker level of -100 $f^{-1}$. The blue shaded area from 10~kHz~-~3~MHz indicates the combined noise of the 1.5~$\mu$m CW laser, FM-to-AM conversion in the optical etalon as well as AM-to-PM conversion in the EOM. We believe the latter to be the dominant noise source in this frequency range. Intensity stabilization of the 10~GHz drive signal could be employed in the future to minimize phase noise due to AM-to-PM in the multiplication setup. From the residual noise measurements above, we find that 10 times opto-electronic multiplication can support the derivation of stable mm-wave signals at a level of 1 part in $10^{16}$ at 1~s averaging.

In principle, the technique of opto-electronic multiplication with filtering can be used to generate microwave/mm-wave signal at any multiple of the EOM drive frequency, up until the cut off of the W-band photodetector~\cite{song2007generation,song2008broadband}. Aside from supporting the very low noise possible with OFD, electro-optic modulation allows for efficient dissemination of microwave and mm-wave signals via a telecom carrier~\cite{qi2005optical}. The latter has immediate applications to radio-over-fiber~\cite{cox2006limits} and at large scale facilities such as free-electron laser centers~\cite{kim2010attosecond}, as well as in radio-astronomy~\cite{cliche2004high,nand2011ultra}. Because opto-electronic multiplication allows for efficient dissemination of microwave and mm-wave signals via a telecom carrier~\cite{qi2005optical}, its combination with very low noise OFD has immediate applications to radio-over-fiber~\cite{cox2006limits} and timing distribution at large scale facilities such as free-electron laser centers~\cite{kim2010attosecond}, and phase-array radio-telescopes~\cite{nand2011ultra,cliche2004high}. Finally, the techniques we introduce should be applicable to the generation of a low-noise microwave and millimeter wave signals with significantly improved stability for improved resolution and precision in optical arbitrary waveform generation, optical signal processing and ranging~\cite{jiang2007optical,li2014photonicweiner}.

\section*{Conclusion}
\begin{figure}
\centering
\includegraphics[width=8cm]{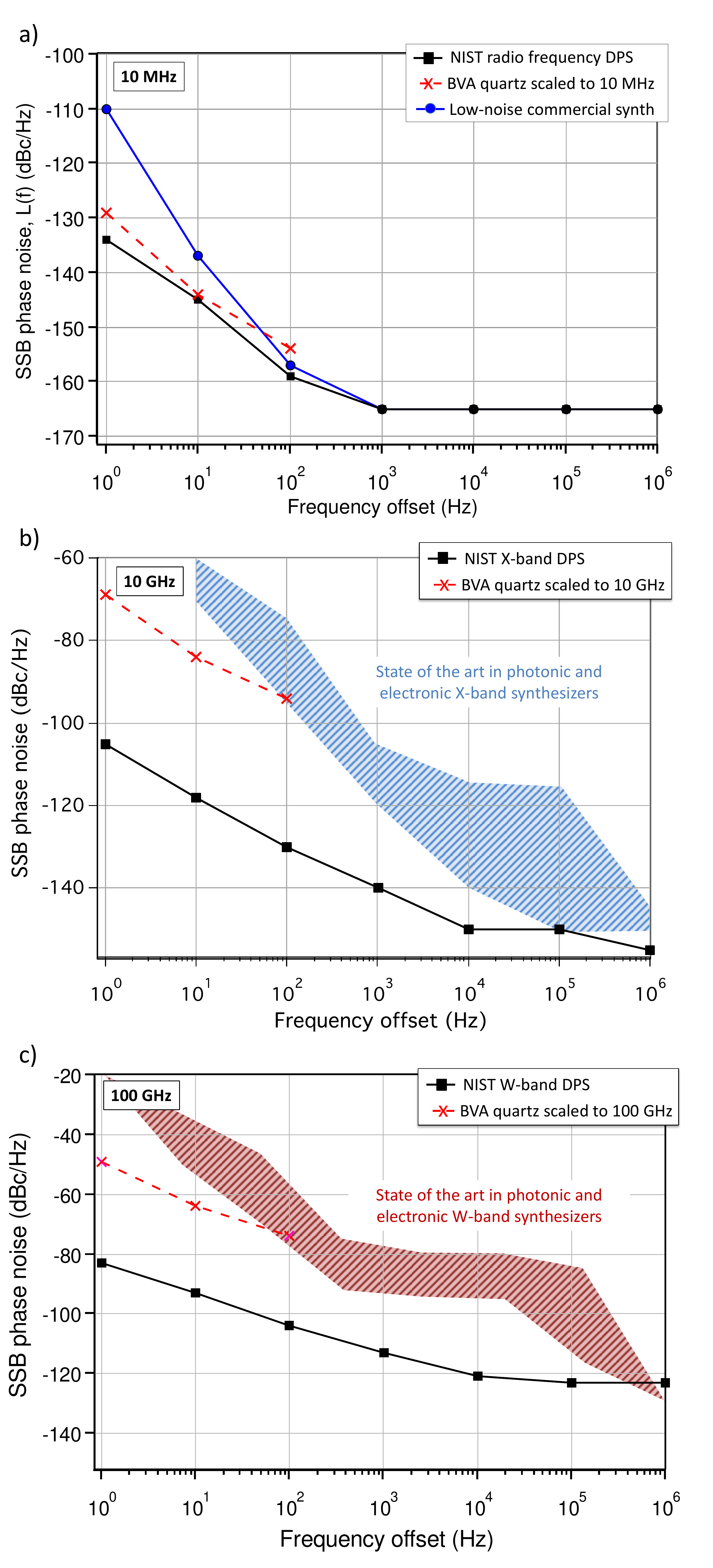}
\caption{Single oscillator comparison of the spectral purity of tunable (shaded region) and fixed sources (dashed line) as compared to DPS for carrier frequencies a) 10~MHz, b) 10~GHz~\cite{eliyahu2013resonant,zhang2014tunable,li2014electro} and c) 100~GHz~\cite{cliche2006precision,pillet2014dual,schneider2013radiofrequency,song2008broadband}.}
\label{figure5}
\end{figure}

We demonstrated ultra low-noise synthesis of agile RF, X-band, and W-band signals using optical frequency division as the time base for direct digital synthesis. Using this technique we have synthesized X-band signals at 10 GHz with a close to carrier noise of -100 dBc/Hz and a fractional frequency stability of close to $1 \times 10^{-15}$. Opto-electronic multiplication provides the extension of synthesis to the W-band at 100 GHz with a close to carrier phase noise level of $-$80 dBc/Hz at 1 Hz, and falling to a noise floor of $-$145 dBc/Hz at 10 MHz. A comparison with other synthesizers and fixed sources in Fig. 5 shows the power of digital-photonic synthesis (DPS). As can be seen in Fig.~\ref{figure5} (a), our DPS demonstrated a spectral purity comparable with the best BVA quartz oscillators but with the benefit of full DDS agility. In the X-band at 10 GHz, we have demonstrated up to 5 orders of magnitude ($>$ 50 dB) improvement in the close-to-carrier phase noise performance for tunable 10~GHz signals (see Fig.~\ref{figure5} (b)). For agile 100~GHz generation, we have demonstrated up to 7 orders of magnitude ($>$70~dB) improvement in the close-in noise, and greater than 20 dB improvement in the high frequency noise floor (see Fig.~\ref{figure5} (c)). Although we limited our measurements to three frequency bands, the technique outlined in this article can allow for a continuously tunable, low-noise source from near DC to greater than 100~GHz. With ongoing developments in compact and robust optical reference cavities~\cite{lee2013spiral} and comb generators~\cite{del2007optical}, DPS with ultra-low noise could move from laboratory demonstrations to system integration in next-generation telecommunications, test and measurement, and radar.

\section*{Methods}
 
In our measurements, characterization of the absolute phase noise and fractional frequency instability was obtained using two nearly identical, but independent digital-photonic synthesizers. This required duplication of the entire digital-photonic setup for comparison against a second system as a reference. In the sections below we describe only a single system, but indicate important differences between systems when relevant.

\noindent
\textbf{Optical reference cavity:}
We employ two optical frequency references, one at 282~THz and the other at 518~THz. The optical frequency references are passive, two-mirror optical cavities constructed with a low-expansion glass ULE spacer with near zero thermal expansion at room temperature. The optical cavity is held in a temperature-controlled vacuum chamber to isolate it from environmental perturbations.  The cavity is mounted such that acceleration of the cavity is minimally coupled to changes in length. The cavity exhibits an impressively small absolute length change that averaged over 1~s is less than the diameter of an atomic nucleus (1~femtometer). An intensity-stabilized, single-frequency laser is locked to a single longitudinal mode of the optical reference cavity using the Pound-Drever Hall stabilization scheme~\cite{drever1983laser}.  The cavity stabilized light is delivered via noise-cancelled optical fiber~\cite{ma1994delivering} to the modelocked laser that performs division of the optical carrier. Ideally, division of the 282~THz (518~THz) optical carrier to 10~GHz or 100~GHz would reduce the optical phase noise power spectral density by 89~dB (92~dB) or 69~dB(72~dB), respectively.

\noindent
\textbf{Optical frequency comb divider:}
In our experiments we used two optical frequency comb dividers based on passively modelocked octave spanning Ti:Sapphire lasers~\cite{fortier2006octave}, both with a native repetition rate of 1 GHz (pulse period of 1~ns, round trip cavity length of 30 cm).  Each laser produced greater than 1~W average power with a usable optical bandwidth from 600$-$1200 nm. The optical comb spectrum is characterized by two RF frequencies, the laser repetition rate $f_{rep}$ and the laser offset frequency $f_0$. Any optical mode, $n$, of the comb can be described as $\nu_n = f_0 + nf_{rep}$. The optical spectrum of an OFD is stabilized by acting on the laser cavity length using a piezo-actuated mirror, and on the laser pump power via an acousto-optic modulator. This allows for control of the laser repetition rate and the laser offset frequency, respectively.  The laser offset frequency is measured using the technique of self-referencing~\cite{jones2000carrier}. The OFD spectrum is stabilized to the optical reference via optical heterodyne with the cavity-stabilized laser described above~\cite{diddams2000direct}. This generates a RF beat note that measures the difference between a single comb line and the cavity-stabilized laser.  Feeding back to the cavity length of the OFD to stabilize the frequency of the optical beat signal against a synthesized RF frequency transfers the stability of the optical reference to every optical mode of the OFD.

\noindent
\textbf{Optical photodetection of the frequency comb:}
Stabilization of the modelocked laser spectrum, in the manner described above, results in an optical pulse train with a pulse-to-pulse timing error of less than 1 femtosecond. Photodetection with a modified-uni-travelling carrier (MUTC) photodiode~\cite{li2010high} is then used to convert the optical pulse train to an electrical one. Pulse interleaving~\cite{haboucha2011optical} that effectively multiplies the pulse repetition rate from 1 GHz to 2 GHz, is used to help to alleviate nonlinearties in photodetection by reducing the energy per pulse.

\noindent
\textbf{Frequency conversion using DDS and OFD:}
We employed a DDS (Analog Devices 9914\footnote{*Any use of trade, firm, or product names is for descriptive purposes only and does not imply endorsement by the U.S. Government.}) that can synthesize analog signals, $f_{DDS}$, up to 800 MHz when clocked with the lowest order harmonic from the OFD at 2 GHz. For efficient shifting of the 10 GHz OFD harmonic by the output of the DDS we employed an IQ mixer. This frequency converter has 3 input ports (I, Q, and LO), and one output port. The LO port is driven at +10 dBm with the 10 GHz from the OFD. Ideally, by driving the I and Q ports in quadrature and with equal amplitude at $f_{DDS}$, the IQ mixer produces one strong tone at 10 GHz + $f_{DDS}$ and suppresses both the LO signal at 10 GHz and the image frequency at 10~GHz$-f_{DDS}$. By switching the phases of the I and Q ports by 90 degrees one can choose which image signal is selected. For simplicity a broadband hybrid coupler (5-500 MHz) was employed to split the signal from the DDS input to the I and Q ports of the mixer, which limited our synthesized X-band bandwidth to 9.5$-$10.5 GHz. Although the amplitude and phase out of the coupler was not perfectly balanced, we observed suppression of the local oscillator and image signals by approximately 30~dB. Careful setting of the phase and amplitude to the I and Q ports of the mixer could significantly improve the suppression of the image frequency, and small DC offsets into the I and Q ports can improve rejection of the LO signal.  

The synthesized X-band signals output from two similar but fully independent X-band synthesizers were compared using a double balanced X-band mixer. In our phase noise measurements, the difference frequency between the two tunable 10 GHz signals was always maintained at a difference frequency of 25 MHz. This intermediate frequency was input to a Symmetricom 5125A* phase noise analyzer for measurement of the phase noise and Allan deviation. A low noise 5 MHz quartz oscillator was used as a stable reference for phase noise and frequency counting measurements ( 5 MHz phase noise of -120~dBc/Hz at 1 Hz offset). 

\noindent
\textbf{Opto-electronic multiplication and characterization of 100 GHz signals:}
For generation and comparison of agile 100 GHz signals, we employed low-loss electro-optic modulators, each driven by independent 10 GHz signals with carrier strengths between 25-30 dBm. The EOMs were used to phase modulate a single 20 mW, 1.5~$\mu$m CW laser. Prior to being split between the two EOMs, the laser was amplified to 200~mW using a polarization maintaining Er-doped fiber amplifier. The modulated CW laser was centered midway between the optical modes of the 100 GHz fiber-coupled optical etalons, each with a 10 GHz optical linewidth. Consequently, the etalon only passed harmonics with the order +5 and -5, yielding two optical lines separated by 100 GHz with residual 10 GHz sidebands -10 dB down from the optical carrier. Using two NEL IOD-PMF*, W-band UTC photodiodes, biased at -2 V and directly coupled to a WR10 waveguide, we were able to extract a 100 GHz carrier of -5 dBm (5 mA photocurrent) with approximately 15 mW incident on each photodetector. 

The tunable 100 GHz signals from two photodetectors were compared using a W-band phase bridge built from waveguide components. The two branches in the phase bridge, LO and RF, took the signals output from two photodiodes and compared them using a W-band mixer. Both branches of the phase bridge required amplification for proper operation of the mixer. A low-noise amplifier was used in the RF branch to drive the RF port of the mixer at 0 dBm and a power amplifier was used to drive the LO port of the mixer at +15 dBm. The intermediate frequency output from the W-band mixer at 50 MHz was filtered and amplified and then input to a Symmetricom phase noise test set to evaluate the phase noise and frequency stability. 

Evaluation of the W-band photodetector and opto-electronic multiplication setup residual noise, as well as measurement of the absolute 100 GHz phase noise to 10 MHz offset, required that the W-band phase bridge was driven in quadrature by two 100 GHz signals of the same frequency. A phase shifter in the RF branch of the phase bridge was used to place the signals in quadrature. The DC signal output from the mixer was then amplified with a low noise IF amplifier and the combined phase noise measured with vector signal analyzer with a bandwidth of 10~MHz.

\section*{Acknowledgement}
The authors would like to acknowledge the DARPA PULSE program for funding. A. Rolland is supported by La D\'{e}l\'{e}gation G\'{e}n\'{e}rale de l'Armement. We would also like to thank C. Nelson, B. Riddle, P. Hale and A. Weiner for helpful discussions and loan of equipment in the initial stages of our research. This work is a contribution of the US Government
and is not subject to copyright in the US.

\bibliographystyle{ieeetr}
\bibliography{references}

\end{document}